\documentclass[aps,prl,twocolumn,showpacs]{revtex4}
\usepackage{graphicx,color}

\begin{document}

\title{Superconductivity in the 2D Hubbard model: 
Electron doping is different}

\author{D.\ Eichenberger and D.\ Baeriswyl}
\affiliation{Department of Physics, University of Fribourg, 
CH-1700 Fribourg, Switzerland.}

\date{\today}

\begin{abstract}

A variational Monte Carlo calculation is used for studying the ground state
of the two-dimensional Hubbard model, including hopping between both nearest 
and next-nearest neighbor sites. Superconductivity with $d$-wave symmetry
is found to be restricted to densities where the Fermi surface crosses
the magnetic zone boundary. The condensate energy is much larger for hole
doping than for electron doping. Superconductivity is kinetic energy driven
for hole doping, but potential energy driven for electron doping. Our findings
agree surprisingly well with experimental data for layered cuprates, 
both for electron- and hole-doped materials.
\end{abstract}


\maketitle

It is widely accepted that the layered cuprates are doped Mott insulators, 
which are well described by the two-dimensional Hubbard model, at least
regarding the insulating antiferromagnetic state of the parent compounds and
the transition to a metallic state upon doping.
Whether this model embodies superconductivity has been debated since two 
decades, but recent
results obtained with improved numerical techniques strengthen the case of 
pairing induced by on-site repulsion. Progress has been made both in 
dynamical mean-field theory
\cite{Maie1,Haul1}, where an effective attraction is deduced from the
irreducible two-particle vertex,  and in variational calculations 
\cite{Yoko2, Eich2}, where a broken-symmetry ground state with a $d$-wave
superconducting order parameter is found in a certain density range
and for large enough values of $U$. 

The Hubbard Hamiltonian 
$\hat{H}=\hat{H_0}+U\hat{D}$ consists of a 
hopping term (``kinetic energy'')
\begin{equation}\label{ansatz}
\hat{H}_0=-\sum_{i,j,\sigma}t_{ij}(c^\dag_{i\sigma}c_{j\sigma}
+c^\dag_{j\sigma}c_{i\sigma})
\nonumber
\end{equation}
and an on-site repulsion $U\hat{D}$, where 
$\hat{D}=\sum_i n_{i\uparrow}n_{i\downarrow}$ is the 
number of doubly occupied sites, $n_{i\sigma}=c_{i\sigma}^\dag c_{i\sigma}$,
and the operator $c_{i\sigma}^\dag$ ($c_{i\sigma}$) creates (annihilates)
an electron at site $i$ with spin $\sigma$.

In our previous work \cite{Eich2} we have restricted ourselves to
nearest-neighbor hopping ($t_{ij}=t$ for nearest-neighbor sites, 
0 otherwise). We have used the variational ansatz
\begin{equation}
\vert\Psi\rangle=e^{h\hat{H}_0/t}e^{-g\hat D}\vert\Psi_{0}\rangle\, ,
\end{equation}
which yields a substantial improvement with respect to the Gutzwiller
wave function ($h=0$), and comes very close to the exact ground state both for small 2D systems \cite{Otsu1} and for the solvable 1/r chain \cite{Dzie1}. Both antiferromagnetic and
superconducting states have been chosen as 
mean-field states $\vert\Psi_0\rangle$. Antiferromagnetism was found 
to prevail at half filling, while $d$-wave pairing has been obtained for the 
doped system, below a hole concentration of 0.18. At first sight these results 
seem to contradict recent work
by Aimi and Imada \cite{Aimi1}, who see no signature for superconductivity in
their Monte Carlo simulation. A closer examination shows that 
there is no discrepancy. On the one hand, the hole densities 
considered by Aimi and Imada for up to 10x10 lattices (0.18, 0.22, 0.28) are 
in a region where we obtained a vanishing superconducting order parameter
\cite{Eich2}.
On the other hand, for the only other density studied by Aimi and Imada (0.09) 
their result is inconclusive. In fact, for their parameter values (U/t=6, 
lattice size 8x8) the order parameter is expected to be too small to lead to
a contradiction \cite{Yoko2}. 

\begin{figure}[!t]
\includegraphics[width=0.3\textwidth]{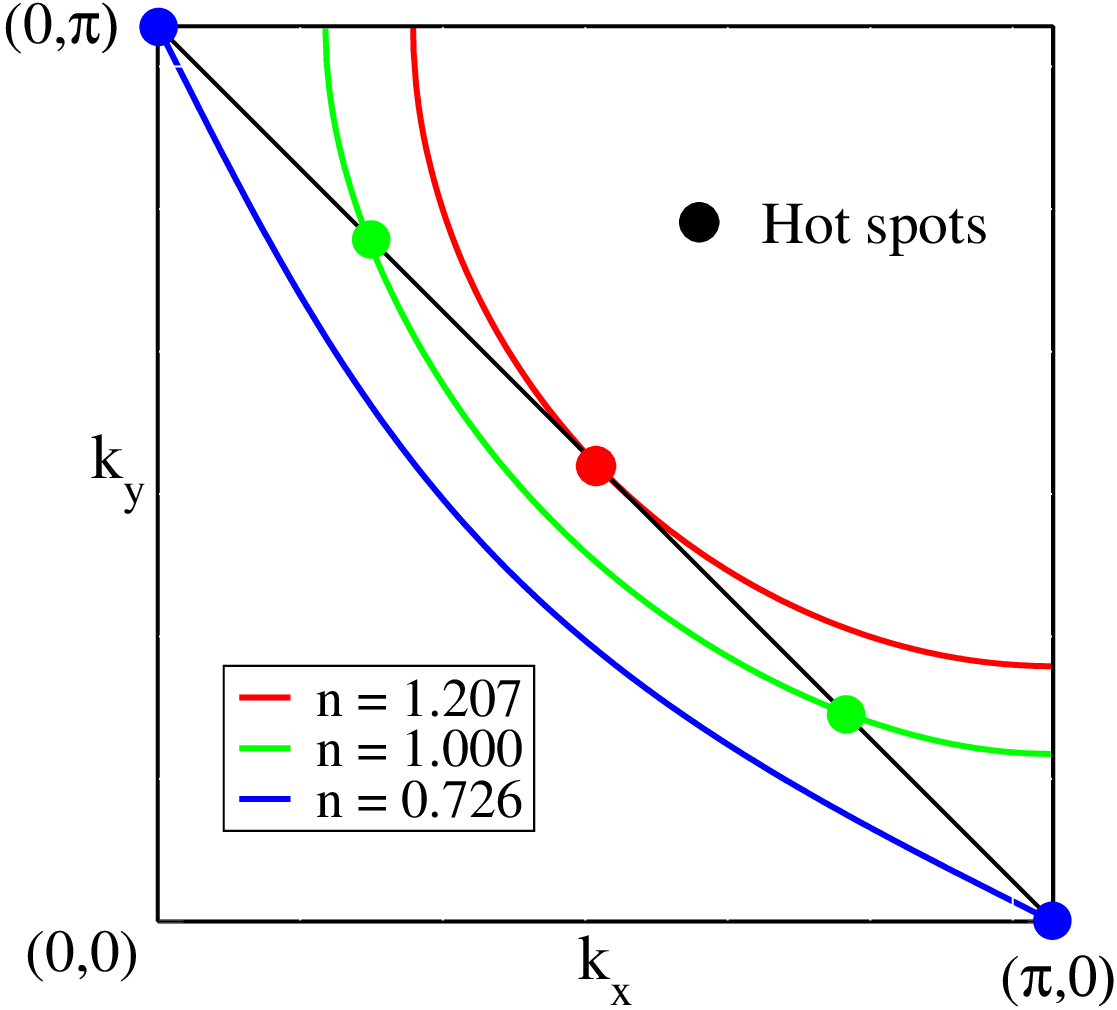}
\caption{\label{FS}(Color online) Fermi surface for three different electron
densities.}
\end{figure} 

With hopping restricted to
nearest-neighbor sites the Hubbard Hamiltonian (on the square lattice)
is electron-hole symmetric and there is no difference between electron and
hole doping. However, this restriction leads to a (bare) Fermi surface
which disagrees qualitatively with photoemission experiments on layered cuprates \cite{Dama1}. In 
the present paper we investigate the more realistic case where hopping between
both nearest ($t$) and next-nearest neighbors ($t'$) is included. We use
parameters $U=8t$ and $t'=-0.3t$ throughout. The bare single-particle
spectrum  
\begin{eqnarray}
\epsilon_{\vec{k}}=-2t(\cos{k_x}+\cos{k_y})-4t'\cos{k_x}\cos{k_y}
\nonumber
\end{eqnarray}
leads to the Fermi surfaces of Fig.\ \ref{FS}. The innermost line corresponds 
to the van Hove filling where
the Fermi surface passes through the saddle points at $(\pi,0)$ and $(0,\pi)$.
These crossings between the Fermi surface and the magnetic zone boundary
(the ``hot spots'') move inwards as the density $n$ is increased and finally
merge (outermost line). Hot spots are restricted to $0.726<n<1.206$. Our
results, to be discussed below, indicate that superconductivity occurs
only in this range.

Here we use again the variational ansatz of Eq.\ (\ref{ansatz}), but restrict
ourselves on a $d$-wave superconducting ground state, which introduces,
in addition to $g$ and $h$, two other parameters, the amplitude $\Delta$ of
the superconducting gap function and the ``chemical potential'' $\mu$.
To compute the variational energy, the exponent of the operator 
$e^{-g\hat{D}}$ is first decoupled by applying a discrete Hubbard-Stratonovich 
transformation \cite{Hirs2}, which introduces an Ising spin at each site. 
All operators are then quadratic in creation and annihilation operators and 
therefore the fermionic degrees of freedom can be integrated out. The 
remaining sum over Ising spin configurations is performed by a Monte Carlo 
simulation. In order to avoid the minus sign problem, calculations are carried 
out in the grand canonical ensemble with an average density fixed by 
$\mu$, which is therefore not a variational parameter.
To reduce the statistical error, the optimization procedure is based on the 
method proposed by Ceperley {\it et al.} \cite{Kalo1, Umri1}. We have used
periodic-antiperiodic boundary conditions.

\begin{center}
\begin{table}
\begin{tabular}{ccccc}
\hline
\hline
n&\hspace{0.4cm}$\mu$\hspace{0.4cm}&\hspace{0.4cm}g\hspace{0.4cm}&\hspace{0.4cm}h\hspace{0.4cm}&\hspace{0.4cm}E/t\hspace{0.4cm}\\
\hline
0.7500&\hspace{0.4cm}-0.9921(1)\hspace{0.4cm}&\hspace{0.4cm}4.2(1)\hspace{0.4cm}&\hspace{0.4cm}0.113(2)\hspace{0.4cm}&\hspace{0.4cm}-0.858(1)\hspace{0.4cm}\\
0.7800&\hspace{0.4cm}-0.9612(3)\hspace{0.4cm}&\hspace{0.4cm}4.0(1)\hspace{0.4cm}&\hspace{0.4cm}0.112(2)\hspace{0.4cm}&\hspace{0.4cm}-0.829(1)\hspace{0.4cm}\\
0.8125&\hspace{0.4cm}-0.9107(3)\hspace{0.4cm}&\hspace{0.4cm}3.9(1)\hspace{0.4cm}&\hspace{0.4cm}0.111(2)\hspace{0.4cm}&\hspace{0.4cm}-0.795(1)\hspace{0.4cm}\\
0.8400&\hspace{0.4cm}-0.788(1)\hspace{0.4cm}&\hspace{0.4cm}3.7(1)\hspace{0.4cm}&\hspace{0.4cm}0.111(2)\hspace{0.4cm}&\hspace{0.4cm}-0.763(1)\hspace{0.4cm}\\
0.9000&\hspace{0.4cm}-0.728(1)\hspace{0.4cm}&\hspace{0.4cm}3.8(1)\hspace{0.4cm}&\hspace{0.4cm}0.111(2)\hspace{0.4cm}&\hspace{0.4cm}-0.676(1)\hspace{0.4cm}\\
0.9500&\hspace{0.4cm}-0.603(1)\hspace{0.4cm}&\hspace{0.4cm}4.0(1)\hspace{0.4cm}&\hspace{0.4cm}0.114(2)\hspace{0.4cm}&\hspace{0.4cm}-0.591(1)\hspace{0.4cm}\\
\hline
\hline
\end{tabular}
\caption{\label{h_dop_tab} Chemical potential, parameters g and h and 
energy per site for hole doping ($n<1$) and an $8\times 8$ lattice.}
\end{table}
\end{center}

\begin{center}
\begin{table}
\begin{tabular}{ccccc}
\hline
\hline
n&\hspace{0.4cm}$\mu$\hspace{0.4cm}&\hspace{0.4cm}g\hspace{0.4cm}&\hspace{0.4cm}h\hspace{0.4cm}&\hspace{0.4cm}E/t\hspace{0.4cm}\\
\hline
1.0500&\hspace{0.4cm}0.7666(1)\hspace{0.4cm}&\hspace{0.4cm}3.6(1)\hspace{0.4cm}&\hspace{0.4cm}0.109(2)\hspace{0.4cm}&\hspace{0.4cm}-0.222(1)\hspace{0.4cm}\\
1.0800&\hspace{0.4cm}0.6440(1)\hspace{0.4cm}&\hspace{0.4cm}3.4(1)\hspace{0.4cm}&\hspace{0.4cm}0.106(2)\hspace{0.4cm}&\hspace{0.4cm}-0.069(1)\hspace{0.4cm}\\
1.1000&\hspace{0.4cm}0.5488(1)\hspace{0.4cm}&\hspace{0.4cm}3.2(1)\hspace{0.4cm}&\hspace{0.4cm}0.104(2)\hspace{0.4cm}&\hspace{0.4cm}0.040(1)\hspace{0.4cm}\\
1.1300&\hspace{0.4cm}0.4380(1)\hspace{0.4cm}&\hspace{0.4cm}3.0(1)\hspace{0.4cm}&\hspace{0.4cm}0.100(2)\hspace{0.4cm}&\hspace{0.4cm}0.206(1)\hspace{0.4cm}\\
1.1600&\hspace{0.4cm}0.3870(1)\hspace{0.4cm}&\hspace{0.4cm}2.9(2)\hspace{0.4cm}&\hspace{0.4cm}0.096(3)\hspace{0.4cm}&\hspace{0.4cm}0.374(1)\hspace{0.4cm}\\
1.2000&\hspace{0.4cm}-0.2996(1)\hspace{0.4cm}&\hspace{0.4cm}2.6(2)\hspace{0.4cm}&\hspace{0.4cm}0.091(3)\hspace{0.4cm}&\hspace{0.4cm}0.608(1)\hspace{0.4cm}\\
\hline
\hline
\end{tabular}
\caption{\label{e_dop_tab} Chemical potential, parameters g and h 
and energy per site for electron doping ($n>1$) and an $8\times 8$ 
lattice.}
\end{table}
\end{center}

\vspace{-2.0cm}

\begin{figure}[!t]
\includegraphics[width=0.5\textwidth]{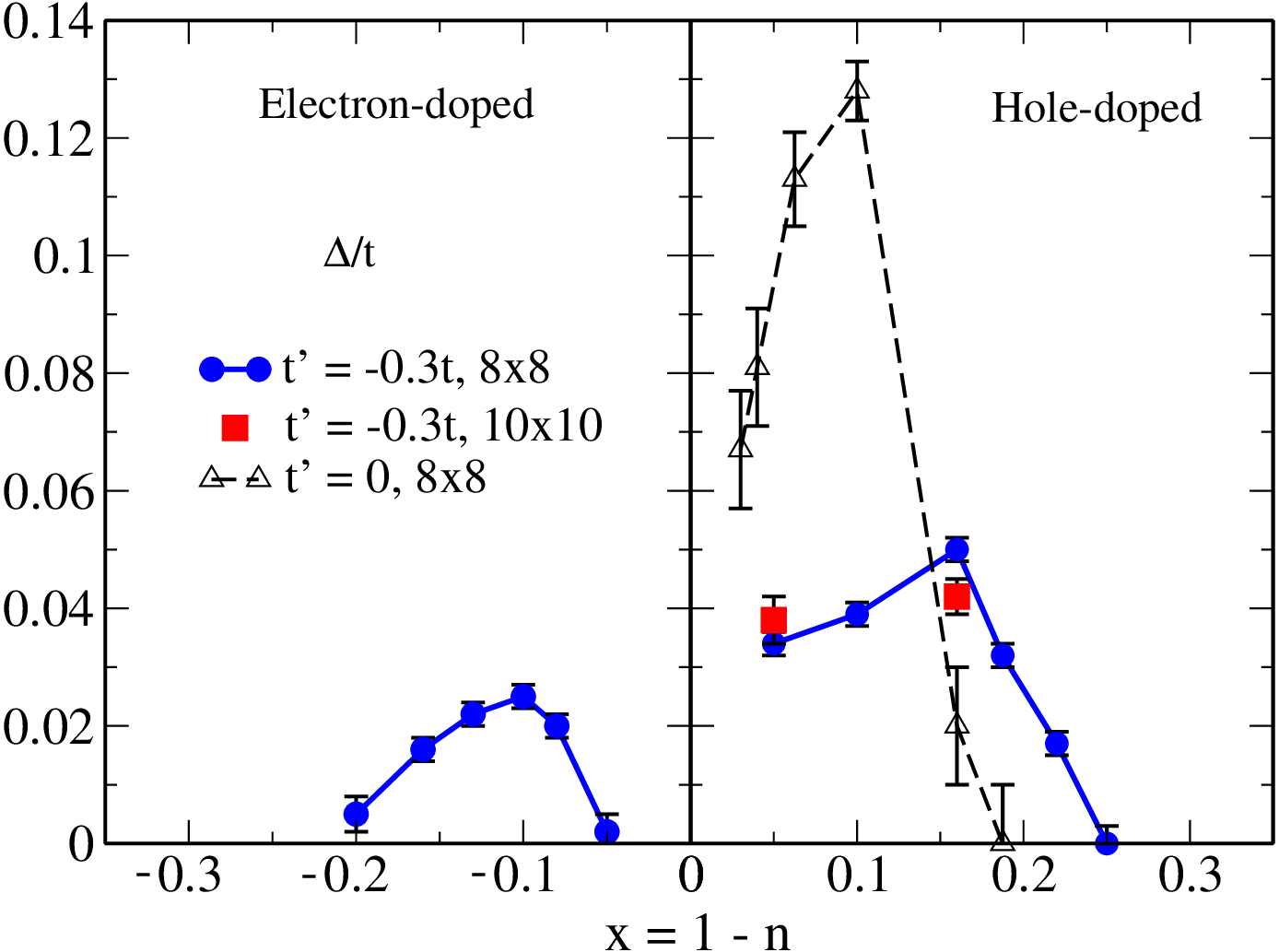}
\caption{\label{Delta}(Color online) Optimized gap parameter as a function of 
doping for two different lattice sizes. For comparison, the result for $t'=0$ 
is also shown (from Ref.\ \cite{Eich2}).}
\end{figure} 

\begin{figure}[!t]
\includegraphics[width=0.5\textwidth]{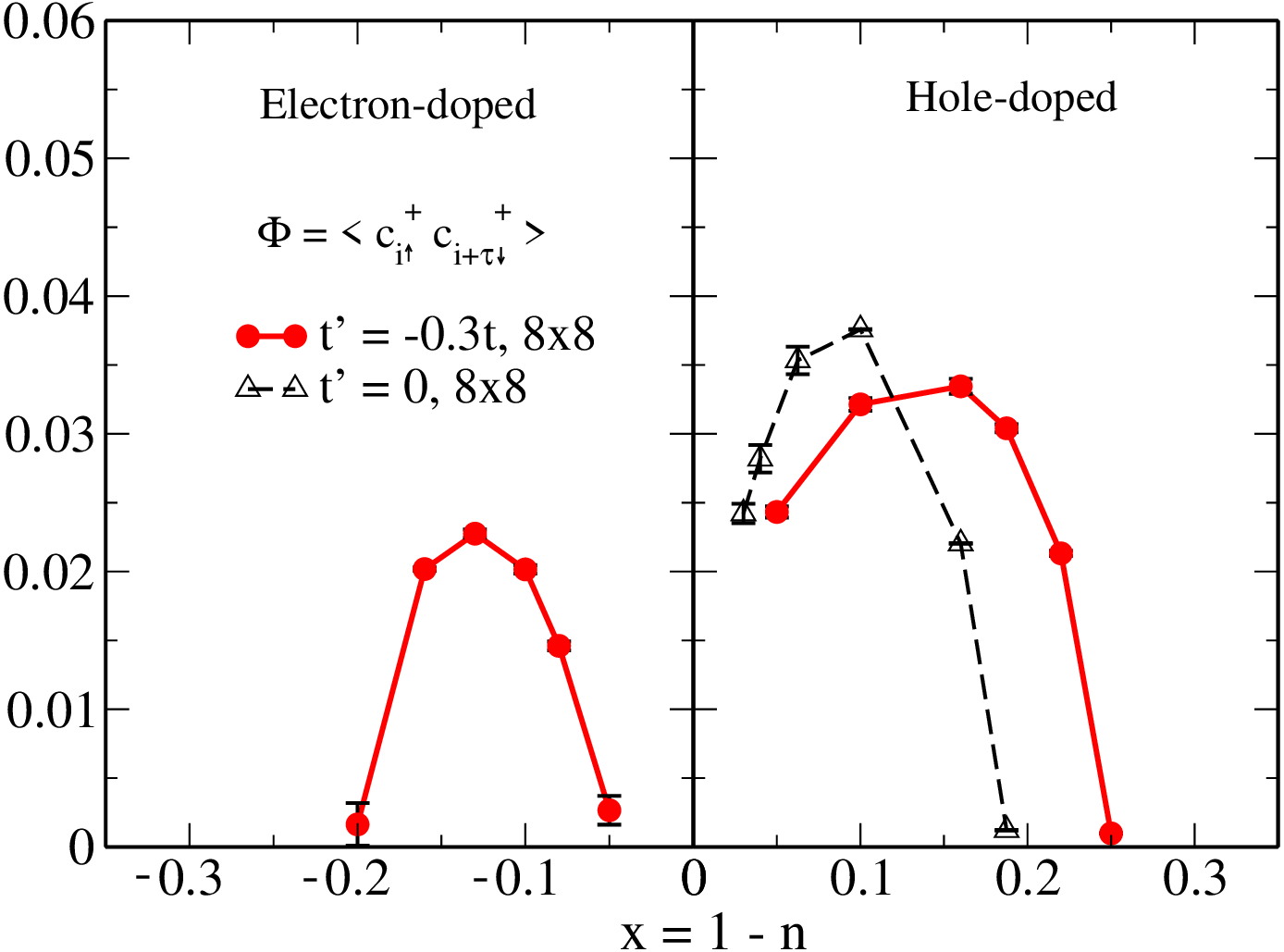}
\caption{\label{PO}(Color online) Superconducting order parameter as a 
function of doping for an $8\times 8$ lattice.}
\end{figure}

\begin{figure}[!t]
\includegraphics[width=0.51\textwidth]{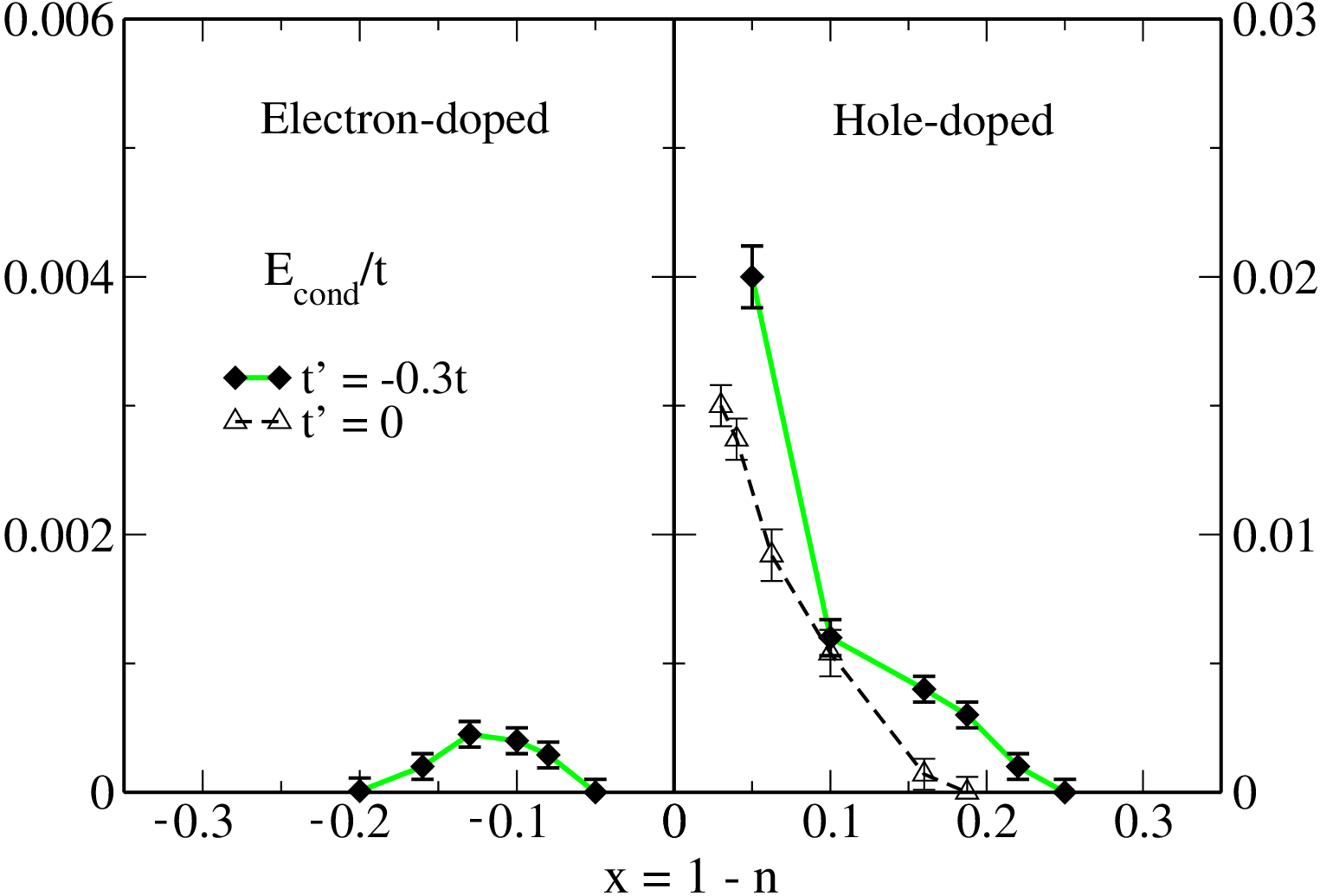}
\caption{\label{Econd}(Color online) Condensation energy per site for an 
$8\times8$ lattice.}
\end{figure}

The minimization of the energy E for fixed average densities yields the
results of Tables I and II for hole and electron doping, respectively. We 
notice that the parameter $g$, which controls the crossover between
itinerant (small $g$) and ``localized'' (large $g$) many-particle states, 
remains large for hole doping, but decreases rapidly for electron doping.
Therefore, while the hole-doped region $0.75<n<0.95$ is a ``localized''
doped Mott insulator, the electron-doped part $1.05<n<1.2$ rapidly undergoes
a crossover to an itinerant regime as $n$ increases. We attribute this
difference to the bare single-particle density of states at the Fermi energy,
which is much larger for hole than for electron doping. 

Figs.\ \ref{Delta} and \ref{PO} show, respectively, the gap parameter 
$\Delta$ and the superconducting order parameter 
$\Phi=\vert\langle c_{i\uparrow}^{\dag}c_{j_{i}\downarrow}^{\dag}\rangle\vert$
as functions of doping concentration $x=1-n$. 
The corresponding results for $t'=0$
\cite{Eich2} are completely electron-hole symmetric and  
reproduced only in the right panels. 
On the hole-doped side superconductivity exists for $0<x<0.25$ 
($\Delta$ remains finite, but $\Phi$ vanishes for $x\rightarrow 0$), $i.e.$
in a larger region than for $t'=0$. In contrast, 
on the electron-doped side the superconducting region is reduced to
$-0.2<x<-0.05$. Thus we find indeed that superconductivity is restricted
to densities where the (bare) Fermi surface passes through hot spots
(see Fig.\ \ref{FS}). The qualitative difference between 
electron- and hole-doping for $x\rightarrow 0$ suggests that superconductivity competes strongly 
with antiferromagnetic order (which we did not take into account in this study) on the hole-doped side and weakly on the electron-doped side. Further
support for this conclusion comes from the condensation energy $E_{cond}=E(0)-E(\Delta)$, which is one 
order of magnitude smaller for electron doping than for hole doping, as 
depicted in Fig.\ \ref{Econd}. 

\begin{figure}[!t]
\includegraphics[width=0.5\textwidth]{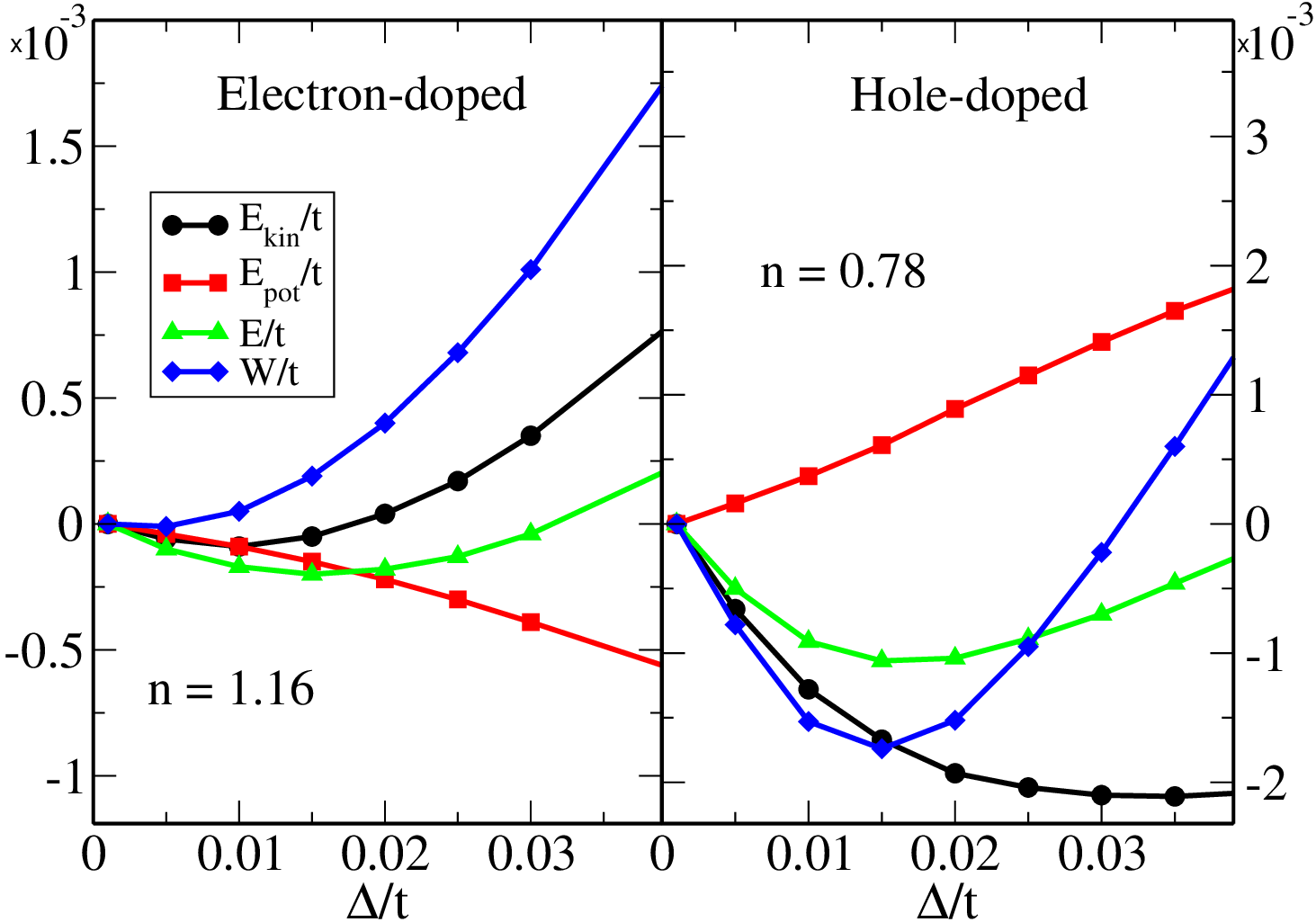}
\caption{\label{EEE}(Color online) Changes in kinetic, potential and total 
energies as well as in the quantity $W$ as functions of the gap parameter, 
for an $8\times8$ lattice. ($-W$ is proportional to the oscillator strength
for intraband transitions.) The relative uncertainties are smaller than the 
symbol sizes.}
\end{figure}

Fig.\ \ref{EEE} shows the kinetic, potential and total energies as functions 
of the gap parameter for the two densities $n=0.78$ and $1.16$. 
For hole doping the energy gain is
clearly due to a decrease in kinetic energy, while for electron doping the 
decrease in potential energy gives the main contribution to the condensation
energy. The same behavior has been consistently obtained for other densities. The quantity 
\begin{equation}
W=-2\sum_{\vec{k}}
\frac{\partial^2\epsilon_{\vec{k}}}{\partial k_x^2}\,
\langle n_{\vec{k}}\rangle\, ,
\end{equation}
also plotted in Fig.\ \ref{EEE}, is -- up to a minus sign -- proportional 
to the integrated optical conductivity originating from intraband transitions 
\cite{Baer2}. For the simple Hubbard model ($t'=0$) $W$ is equal to the 
kinetic energy, but for $t'\neq 0$ the two quantities differ. For hole doping 
$W$ has a pronounced minimum in the region of the optimal gap,
corresponding to an increase of oscillator strength, while for electron 
doping $W$ increases monotonically with $\Delta$, akin to BCS behavior where 
the oscillator strength is reduced at the onset of superconductivity.

We attribute this asymmetry between hole and electron doping to the
different values of the correlation parameter $g$ (see Tables \ref{h_dop_tab}
and \ref{e_dop_tab}), which puts the hole-doped system into the
``localized'' regime, while the electron-doped system is more itinerant.
To make the point clear we consider the simple 
Hubbard model ($t'=0$) both in the small $U$ (itinerant) and large $U$
(``localized'') limits. In the small $U$ limit superconductivity is produced 
by the Kohn-Luttinger mechanism \cite{Kohn1}, where the condensation energy
is expected to increase as a function of $U/t$, whereas in the large $U$ limit
the condensation energy arises from magnetic exchange and thus is likely to 
increase with $t/U$. The change in kinetic energy is then obtained through
the Hellman-Feynman theorem,
\begin{equation}
E_{kin}(\Delta)-E_{kin}(0)=-t\frac{\partial}{\partial t}E_{cond}\, ,
\end{equation}
which is positive in the small $U$ limit and negative in the large $U$ limit.

\begin{figure}[!t]
\includegraphics[width=0.5\textwidth]{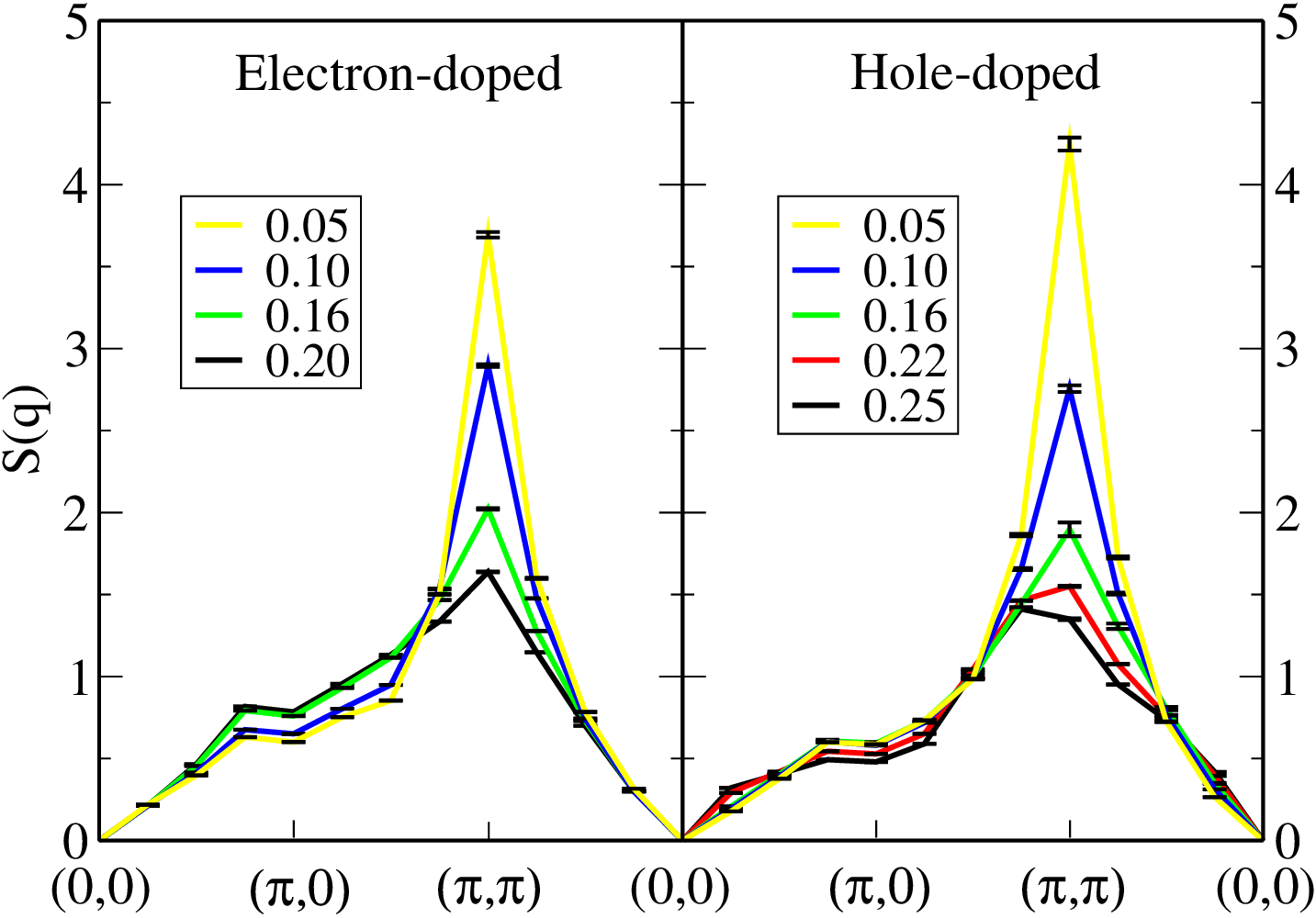}
\caption{\label{SM}(Color online) Magnetic structure factor as a function of the momentum, for various densities above and below half filling, for an $8\times8$ lattice.}
\end{figure}

Additional information can be gained from the magnetic structure factor
\begin{equation}
S(\vec{q})=\frac{1}{N}\sum_{i,j}e^{i\vec{q}\cdot 
(\vec{R}_{i}-\vec{R}_{j})}\langle(n_{i\uparrow}-n_{i\downarrow})
(n_{j\uparrow}-n_{j\downarrow})\rangle\, , 
\end{equation}
displayed in Fig.\ \ref{SM}. $S(\vec{q})$ exhibits a clear maximum at 
$(\pi,\pi)$, which is largest close to half filling and decreases as doping 
increases. There is very little difference between electron and hole doping,
presumably because in the large $U$ limit spin correlations depend mostly
through the exchange constants $J=4t^2/U$ and $J'=4t'^2/U$ on the microscopic
parameters and therefore are essentially independent on the sign of $t'$.

We have seen above that superconductivity is restricted to the region 
where two points of the Fermi surface can be connected (at least approximately)
by the antiferromagnetic wave vector $(\pi,\pi)$. This together with the 
strong peak of $S(\vec{q})$ for $\vec{q}=(\pi,\pi)$ supports the point of view 
that superconductivity in the two-dimensional Hubbard model is due to a 
magnetic mechanism. This conclusion is now widely accepted,
but the question whether the mechanism is of the RVB type 
\cite{Ande1} or arises from the exchange of spin fluctuations 
\cite{Scal2} 
is presently under debate \cite{Ande4, Maie3}. We cannot solve this problem 
on the basis of our variational calculations, but the distinction between
``localized'' and itinerant regimes
can give additional useful insight. According to our results the former
is appropriate for the hole-doped region, where superconductivity is 
associated with a decrease in kinetic 
energy, the latter regime is found for electron doping, where superconductivity
arises from a more conventional gain in potential energy.

The results described above compare surprisingly well with experiments on 
cuprates, better than our earlier calculations \cite{Eich2}
where only nearest neighbor 
hopping has been taken into account. This concerns the phase diagram, in
particular for electron-doped materials for which a recent systematic study
\cite{Kroc1}
gives a dome-shape (for $T_c$) strikingly similar to our Fig.\ \ref{Delta}.
Photoemission data give values of the superconducting gap $\Delta$ 
in the range $10$-$20$ meV for hole-doped 
compounds (LSCO \cite{Yama2} or Bi2201 \cite{Dama1}) and $\sim 5$ meV for 
an electron-doped compound (NdCeCO \cite{Sato1}). Choosing $t=300$ meV 
(neutron data), our maximum gap parameters are $15$ and $7$ meV for hole
and electron doping, respectively. An increase of oscillator strength, 
predicted by our calculations for hole doping, has been reported on the
basis of optical spectroscopy for an underdoped sample \cite{Carb1}, but
the situation is less clear on the overdoped side. We are not aware of 
corresponding measurements on electron-doped materials. 

In conclusion, our variational search for $d$-wave superconductivity in the
two-dimensional Hubbard model gives a differentiated picture, namely a large 
(moderate) correlation parameter for hole (electron) doping, a gain in
kinetic (potential) energy due to pairing and a large (small) condensation
energy. Our wave function is expected to be better suited for describing
the electron-doped region, which is less ``localized'' than the hole-doped 
region. More work is needed to improve our understanding of the pseudogap 
phase observed in the cuprates for weak hole doping, by studying possible 
competing instabilities.  

We are grateful for financial support from the Swiss National Science 
Foundation through the National Center of Competence in Research 
``Materials with Novel Electronic Properties-MaNEP''. The computations have been partially performed on the Pl\'eiades cluster of the EPFL.

\bibliography{list2}



\end{document}